\documentclass[aps,prl,twocolumn,superscriptaddress]{revtex4-1}
\pdfoutput=1
\usepackage{graphicx}
\usepackage{amsmath,amsfonts,amssymb}
\usepackage[colorlinks=true,linkcolor=blue,citecolor=blue]{hyperref}
\usepackage{siunitx}

\begin{document}
\title{Fabry-P\'{e}rot Resonances in a Graphene/hBN Moir\'{e} Superlattice}

\author{Clevin Handschin}
\affiliation{Department of Physics, University of Basel, Klingelbergstrasse 82, CH-4056 Basel, Switzerland}

\author{P\'{e}ter Makk}
\email{Peter.Makk@unibas.ch}
\affiliation{Department of Physics, University of Basel, Klingelbergstrasse 82, CH-4056 Basel, Switzerland}

\author{Peter Rickhaus}
\affiliation{Department of Physics, University of Basel, Klingelbergstrasse 82, CH-4056 Basel, Switzerland}

\author{Ming-Hao Liu}
\affiliation{Institut f\"{u}r Theoretische Physik, Universit\"{a}t Regensburg, D-93040 Regensburg, Germany}

\author{K. Watanabe}
\affiliation{National Institute for Material Science, 1-1 Namiki, Tsukuba, 305-0044, Japan\\}

\author{T. Taniguchi}
\affiliation{National Institute for Material Science, 1-1 Namiki, Tsukuba, 305-0044, Japan\\}

\author{Klaus Richter}
\affiliation{Institut f\"{u}r Theoretische Physik, Universit\"{a}t Regensburg, D-93040 Regensburg, Germany}

\author{Christian Sch\"onenberger}
\affiliation{Department of Physics, University of Basel, Klingelbergstrasse 82, CH-4056 Basel, Switzerland}

\begin{abstract}
While Fabry-P\'{e}rot (FP) resonances and Moir\'{e} superlattices are intensively studied in graphene on hexagonal boron nitride (hBN), the two effects have not been discussed in their coexistence. Here we investigate the FP oscillations in a ballistic \textit{pnp}-junctions in the presence and absence of a Moir\'{e} superlattice. First, we address the effect of the smoothness of the confining potential on the visibility of the FP resonances and carefully map the evolution of the FP cavity size as a function of densities inside and outside the cavity in the absence of a superlattice, when the cavity is bound by regular \textit{pn}-junctions. Using a sample with a Moir\'{e} superlattice, we next show that an FP cavity can also be formed by interfaces that mimic a \textit{pn}-junction, but are defined through a satellite Dirac point due to the superlattice. We carefully analyse the FP resonances, which can provide insight into the band-reconstruction due to the superlattice.
\end{abstract}

\maketitle

Clean graphene has shown to be an excellent platform to investigate various electron optical experiments, ranging from Fabry-P\'{e}rot (FP) resonances \cite{Campos12b,Grushina13,Rickhaus13,Varlet14,Shalom16,Calado15}, snake states \cite{Rickhaus15,Taychatanapat14}, electron guiding \cite{Rickhaus15_2}, magnetic focusing \cite{Taychatanapat13,Calado14}, Veselago lensing or angle-dependent transmission studies of negative refraction \cite{Lee15,Chen16}. The \textit{pn}-junctions are basic building blocks of many of these experiments, which for example are used to confine electrons into FP cavities, where the \textit{pn}-junctions play the role of semireflective  mirrors. The study of FP resonances have already revealed themselves as a powerful tool to investigate various aspects of graphene. Examples are the $\pi$-shift of the FP resonances at low magnetic field originating from Klein tunnelling \cite{Klein29} in single layer graphene \cite{Shytov08,Young09,Varlet16} or the FP resonances in ultraclean, suspended devices that proved ballistic transport over several microns \cite{Rickhaus13, Grushina13}.

Encapsulation of graphene in hexagonal boron nitride (hBN) \cite{Dean10} can not only allow for quasi-ballistic transport (the mean free path is comparable or larger than the device geometry), but can also be used for the creation of a superlattice in graphene due to the periodic potential modulation of the Moir\'{e} pattern. Such a pattern stems from a small lattice mismtach of graphene and hBN.
The modified bandstructure of graphene, which includes the emergence of satellite Dirac peaks (satellite DPs), is experimentally  observable only for small misalignment angles ($\theta$) between hBN and graphene \cite{Woods14,Yankowitz12}. The formation of a Moir\'{e} superlattice has been demonstrated in various spectroscopy \cite{Yankowitz12,Yu14,Shi14}
and transport \cite{Hunt13,Ponomarenko13,Dean13,Gorbachev14,L_Wang15,Wang15,Greenaway15,Kumar16,Lee16} experiments. Besides experiments, considerable effort has been invested to calculate the band-structure of graphene under the influence of such a Moir\'{e} superlattice \cite{Wallbank13,Moon14,Wallbank15}.

Here we demonstrate confinement using band engineering based on locally gated Moir\'{e} superlattices and the appearance of FP resonances defined by the main and satellite DPs playing the role of reflective barriers. Although several aspects of FP cavities have been investigated, such as the effect of the \textit{pn}-junction smoothness on the visibility of the FP resonances \cite{Rickhaus13}, the electronic tunability of the cavity size has not been studied. First, we will discuss samples with a large misalignment angle between the hBN and graphene. We find that the cavity size is strongly dependent on carrier density in and outside the cavity. Then, we turn to aligned samples with a superlattice structure and take advantage of the large tunability of the cavity size to obtain additional information on the modulated Moir\'{e} band-structure which is yet not fully understood. Our findings are consistent with the electron-hole symmetry breaking of the Moir\'{e} superlattice \cite{Wallbank13}.



\begin{figure*}[htbp!]
    \centering
      \includegraphics[width=\textwidth]{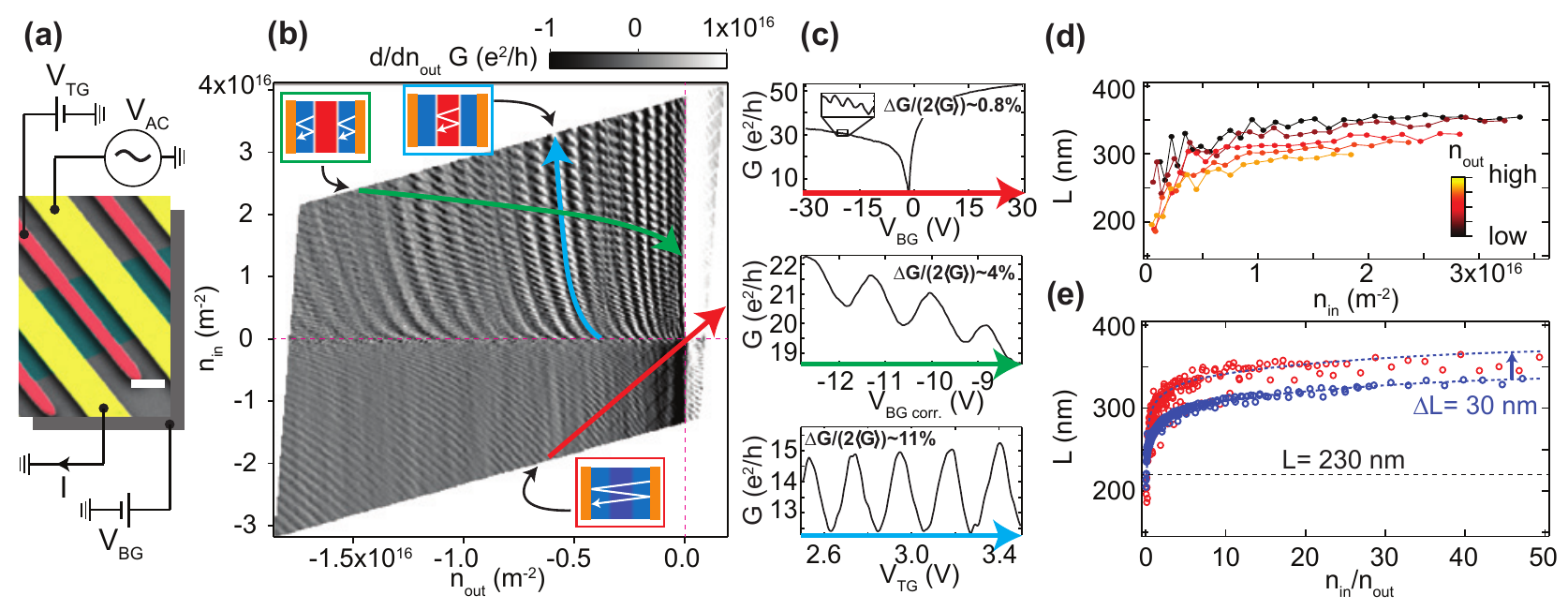}
    \caption{\textbf{Fabry-P\'{e}rot resonances measured in a two-terminal \textit{pnp}-configuration. a,} False-color SEM image with the measurement configuration sketched. The contacts are shown in yellow, the top-gate in red and the graphene between the two hBN-layers is shown in cyan. Scale-bar equals \SI{500}{nm}. \textbf{b,} Numerical derivative of the conductance as a function of global back-gate ($V_{\text{BG}}$) and local top-gate ($V_{\text{TG}}$). Here it is replotted as a function of charge-carrier density in the outer ($n_{\text{out}}$) and inner cavity ($n_{\text{in}}$). The three most important FP resonances present in the system are indicated. \textbf{c,} Visibilty $\Delta G/\left<2G\right>$ of the FP resonances indicated in (b) in comparison. \textbf{d,} Extracted cavity length of the central cavity as a function of $n_{\text{in}}$, as indicated with the cyan arrow in (b), for different values of $n_{\text{out}}$. \textbf{e,} The same data as shown in (d), but now plotted as a function of $n_{\text{in}}/n_{\text{out}}$ for experiment (red) and theory (blue, for transport simulation see Supporting Information). The black, dashed line corresponds to the width of the top-gate measured with the SEM. The blue, dashed lines are a guide to the eye for the theoretical data and the theoretical data shifted by \SI{30}{nm} respectively.}
    \label{fig:Regular_PNP}
\end{figure*}

The device geometry of our two-terminal \textit{pnp}-junction is shown in Fig.~\ref{fig:Regular_PNP}a: a global back-gate and a local top-gate is used to separately control the charge carrier density in the inner ($n_{\text{in}}$) and outer ($n_{\text{out}}$) regions thus forming \textit{pp'p}- or \textit{pnp}-junctions. In order to reduce the cross-talk between the gates we used a thin (\SI{\sim 30}{nm}) hBN layer. Therefore $n_{\text{out}}$ can be calculated according to $n_{\text{out}}=V_{\text{BG}} \cdot (1/C_{\text{SiO}_{2}}+1/C_{\text{hBN,b}})^{-1}$, where $C_{\text{SiO}_{2}}$ and $C_{\text{hBN,b}}$ are the geometrical capacitances of the SiO$_{2}$ and bottom hBN respectively, and $V_{\text{BG}}$ and $V_{\text{TG}}$ the applied gate-voltages. In analogy, $n_{\text{in}}$ is given by $n_{\text{in}}= V_{\text{BG}} \cdot (1/C_{\text{SiO}_{2}}+1/C_{\text{hBN,b}})^{-1}+V_{\text{TG}} \cdot C_{\text{hBN,t}}$ where $C_{\text{hBN,t}}$ is the top hBN capacitance. The Cr/Au contacts dope the graphene in its proximity \textit{n}-type, independent of $V_{\text{BG}}$. The conductance is measured in a two-terminal configuration as a function of $V_{\text{BG}}$ and $V_{\text{TG}}$ (which can be directly converted to $n_{\text{in}}$ and $n_{\text{out}}$), as shown in Fig.~\ref{fig:Regular_PNP}b. The overall device length is roughly \SI{1}{\micro m}, with a top-gate width of \SI{230}{nm} being centered in the middle of the device. In the measured device, a high mobility of the electrons and holes ($\mu_{\text{e,h}}$) was extracted from field effect measurements, yielding $\mu_{\text{e}}$\SI{\sim 150000}{cm^2 V^{-1} s^{-1}} and $\mu_{\text{h}}$\SI{\sim 50000}{cm^2 V^{-1} s^{-1}} respectively.

All structures were fabricated following the procedure given by Wang \textit{et al.} \cite{Wang13} with some additional improvements.  A detailed description of the fabrication is given in the Supporting Information. The measurements were done using low frequency lock-in technique in a variable temperature insert (VTI) with a base-temperature around $T=1.6\,$K.

The derivative of the differential conductance with respect to $n_{\text{in}}$ and $n_{\text{out}}$ are shown in Fig.~\ref{fig:Regular_PNP}b  for $n_{\text{out}}<0$. The visible fringes are FP resonances appearing in the \textit{pp'p} regime (bottom part, $n_{\text{in}}<0$) and \textit{pnp} regime (upper part, $n_{\text{in}}>0$). Specific cuts within an extended (red) or limited (blue, green) gate-range are shown in Fig.~\ref{fig:Regular_PNP}c. While the FP resonances in the inner (outer) cavities are tuned predominantly by $n_{\text{in}}$ ($n_{\text{out}}$) as indicated with the blue (green) arrow in Fig.~\ref{fig:Regular_PNP}b, the FP resonances between the contacts depends on both densities as indicated by the red arrow.
The three types of FP resonances are very different in their visibility  $\Delta G/(2 \left<G\right>)$. Here $\Delta G$ is the difference between the conductance at constructive and destructive interference, and $\left<G\right>$ denotes the mean conduction in between oscillation maximum and minimum. The FP resonances between the contacts yield the lowest visibility ($\sim1\%$), those in the outer cavities yield an intermediate visibility ($\sim4\%$) and FP resonances in the inner cavity yield the highest visibility ($\sim11\%$).
In ballistic graphene, the FP visibilities depend on the transmission/reflection properties of the confining boundaries, namely the \textit{pn}-interface. The transmission/reflection probabilities of a \textit{pn}-interface
strongly depends on the angle of the incoming charge carrier \cite{Cheianov06}.
In the case of a ``sharp" \textit{pn}-junctions ($d\ll\lambda_{F}$, where $d$ denotes the distance over which the density changes and $\lambda_{F}$ is the Fermi wavelength), transmission of charge carriers is possible up to large angles measured with respect to the \textit{pn}-junction normal. In contrast, for a very ``smooth" \textit{pn}-junction ($d\gg\lambda_{F}$), only charge carriers at low angles are transmitted (in our device this is around \SI{\sim 20}{\degree}, although depends strongly on the gate voltages). Since our conductance measurement is not angle resolved, the measured signal averages over all possible angles and leads to smearing of the FP resonances. It turns out that for a smooth \textit{pn}-junction, which transmits a narrower range of angles, the highest FP visibilities can be observed \cite{Rickhaus13}.

In our system we discriminate between two types of \textit{pn}-junctions: i) the \textit{pn}-junction created in proximity of the contacts for $n_{\text{out}}<0$ and ii) the \textit{pn}-junction created using the global back-gate and local top-gate. We note that (ii) is much smoother compared to (i) (see Supporting Information). Following the above given argument one sees that the FP resonance visibility is highest/lowest when the cavity is defined by two softer/sharper \textit{pn}-junctions.

We will now extract the effective cavity length, $L$ from the FP oscillations in the central cavity, which deviates in most cases from the physical width of the top-gate. Assuming a FP resonator with a hard-wall potential (fixed width of the cavity), the cavity width can be extracted from the FP oscillations.
Constructive interference forms if the path-difference between directly transmitted and twice reflected waves is equal to $2 L=j\lambda_{\text{F}}$, where $\lambda_{\text{F}}$ is the Fermi wavelength and $j$ is an integer.
The $j$-th FP resonance can be rewritten as $L\sqrt{n_{\text{j}}}=j\sqrt{\pi}$ using $\lambda_{\text{F}}=2\pi/ k_{\text{F}}=2 \sqrt{\pi/n}$ which is valid for single layer graphene. For two neighboring peaks, for example $j$-th peak at density $n_\text{j}$ and $(j+1)$-th peak at density $n_{\text{j+1}}$:

\begin{equation}\label{cavity-length}
L=\frac{\sqrt{\pi}}{\sqrt{n_{\text{j+1}}}-\sqrt{n_{\text{j}}}} .
\end{equation}

We note here that in Equation~\ref{cavity-length} $L$ is independent of $n_{\text{out}}$ which is an oversimplification of the problem.
An alternative way to define the cavity length is to measure the distance between the two zero-density points of the right and left \textit{pn}-junctions. However, since the position of the \textit{pn}-interface is experimentally not directly accessible we will use Equation~\ref{cavity-length} to deduce the cavity size. More discussion on this is given in the Supporting Information.
The cavity length was extracted by taking various linecuts comparable to the one indicated with the blue arrow in Fig.~\ref{cavity-length}b and then using Equation~\ref{cavity-length}. We observe an increase of $L$ with increasing $n_{\text{in}}$ (for fixed $n_{\text{out}}$), and a decrease of $L$ with increasing $n_{\text{out}}$ (for fixed $n_{\text{in}}$) as shown in Fig.~\ref{fig:Regular_PNP}d. Consequently $L$ does depend on $n_{\text{out}}$, as expected. Surprisingly, by plotting $L$  as a function of  $n_{\text{in}}/n_{\text{out}}$, all data-points lie on one universal curve which is shown in Fig.~\ref{cavity-length}e, independent of the exact position within Fig.~\ref{fig:Regular_PNP}b from which they have been extracted from. Within the applied gate-range, $L$ varies substantially, by up to \SI{200}{nm}, which corresponds to a shift of around \SI{100}{nm} per \textit{pn}-junction.
The evolution of $L$ as a function of $n_{\text{in}}/n_{\text{out}}$ extracted from the experiment was compared with  the one extracted from a transport simulation (based on the method described in Ref. \cite{Rickhaus13}) using Equation~\ref{cavity-length}. The latter reveals good qualitative agreement with the experiment as shown in Fig.~\ref{cavity-length}e. The most significant difference between experiment and theory are: i) an off-set of $\Delta L \sim 30\;$nm from the theory to the experiment and ii) a disagreement between the trends of the two curves when approaching very low charge carrier densities in the outer cavity, corresponding to large values of $n_{\text{in}}/n_{\text{out}}$ (the same is true when depleting the inner cavity, not shown here). In the experiment the cavity length $L$ saturates while it increases continuously in the theory. We believe that the reason for this is that the measured sample bears a residual doping ($n^{*}$) which is not present in theory. The residual doping causes that for values below $n^{*}$ the electrostatic gates are unable to further deplete the graphene, thus for $V_{\text{BG}} \rightarrow 0$ the effective value of $n_{\text{in}}/n_{\text{out}}$ (tuning $L$) remains fixed. The off-set of the two curves (i) might origin from a too narrow top-gate in theory if the top-gate in the experiment was measured with an error of \SI{\sim 30}{nm}.

\begin{figure*}[htbp!]
    \centering
      \includegraphics[width=\textwidth]{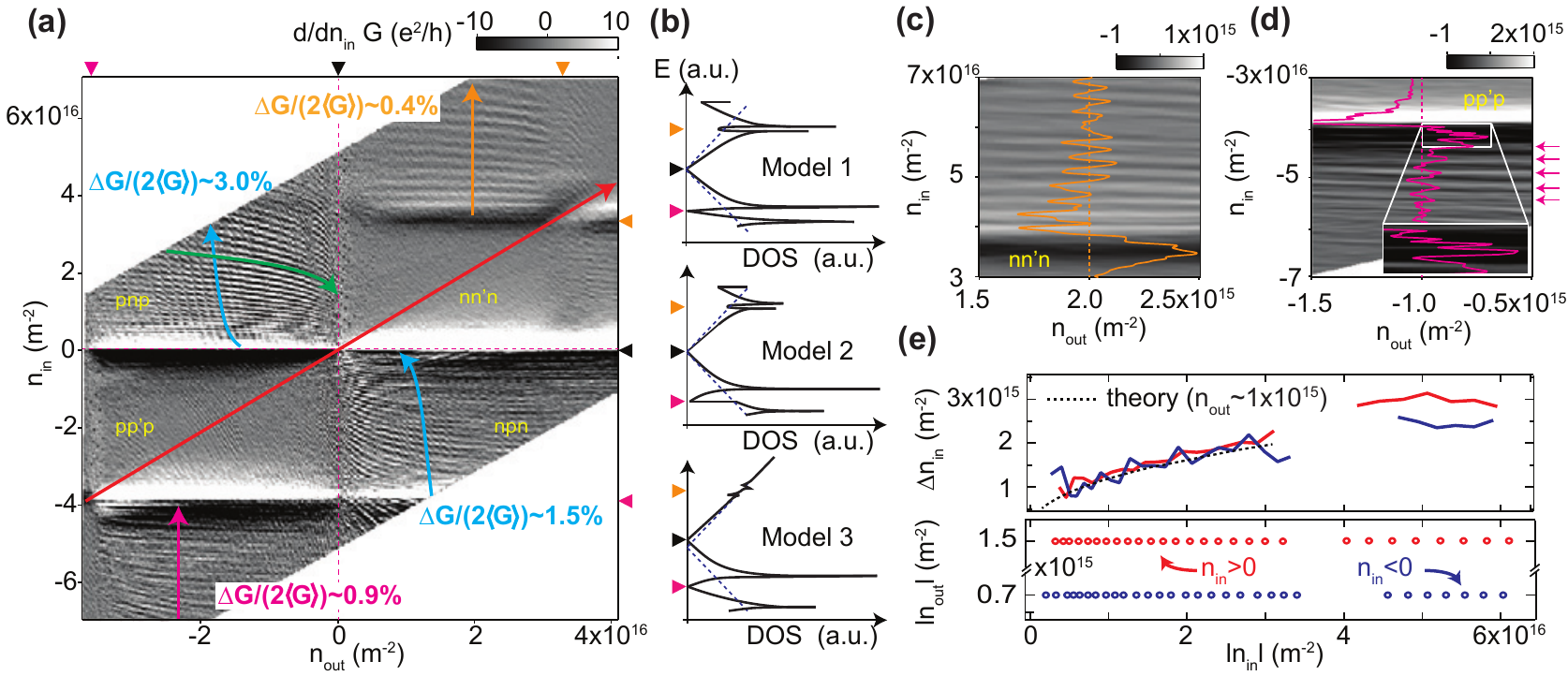}
    \caption{\textbf{Fabry-P\'{e}rot resonancesin in the presence of a Moir\'{e} superlattice. a,} Numerical derivative of the conductance as a function of $n_{\text{in}}$ and $n_{\text{out}}$. The red, cyan and green lines indicate the regular FP resonances described in Fig.~\ref{fig:Regular_PNP}. Additional FP resonances emerge if $n_{\text{in}}$ or $n_{\text{out}}$ is tuned above or below a satellite Dirac peak. These FP resonances are indicated with orange and purple arrows. The position of the main and satellite DPs are indicated with black and purple/orange triangles respectively. \textbf{b,} DOS of graphene as a function of energy in the presence of a Moir\'{e} superlattice with hBN for three different parameter sets used in the calculation performed by Wallbank \textit{et al.} Figure adapted from Ref. \cite{Wallbank13}. Black, orange and purple triangles indicate main and satellite Dirac-peaks respectively. The blue, dashed line indicates the DOS for unperturbed graphene. \textbf{c,d,} High-resolution measurements ($d/dn_{\text{in}}G\;$(e$^2$/h)) of the regions where the inner cavity is tuned beyond the satellite DP and the additional FP resonances are present. \textbf{e,} Position of the individual FP resonance peaks (lower panel) and their relative spacing $\Delta n_{\text{in}}$ (upper panel) is plotted as a function of $n_{\text{in}}$. The black, dashed line indicates the values expected from theory. If the semitransparent boundaries are defined via the satellite DP the spacing is further increased. However, no precise trend of $\Delta n_{\text{in}}$ (increasing or decreasing) as a function of $n_{\text{in}}$ can be seen from the few data-points extracted.}
    \label{fig:Moiree_PNP}
\end{figure*}

Now we turn to graphene which is aligned with a small twist angle with respect to one of the hBN layers. The resulting band-reconstruction includes additional satellite DPs which are indicated with the purple and orange triangles in Fig.~\ref{fig:Moiree_PNP}a where the derivative of the conductance is plotted. The main DP is indicated with the black triangle. At low doping the semitransparent boundaries in the bipolar region are defined by the main DP (comparable to Fig.~\ref{fig:Regular_PNP}b) and FP resonances within inner and outer cavity are observed, marked by the green and cyan arrows, respectively.

Whereas in Fig.~\ref{fig:Regular_PNP}b,c the contacts are the only boundaries leading to FP resonances in the unipolar regime (red arrow), additional semitransparent boundaries, formed by the satellite DPs, emerge at high doping in the presence of a superlattice. As a result, novel FP oscillation are visible at high doping as indicated with the purple and orange arrows in Fig.~\ref{fig:Moiree_PNP}a, which are absent in Fig.~\ref{fig:Regular_PNP}b (no Moir\'{e} superlattice). This new set of  FP oscillation resembles again the pattern known from the bipolar regime and the transition from FP resonances across the whole sample to FP resonances within the inner and outer cavity is a direct consequence of the satellite DPs forming these additional semitransparent boundaries. There are also charge carriers that bounce between the contacts and the satellite DP boundary, leading to weak resonances as a function of $n_{\text{out}}$ (see Supporting Information).


In the following, we compare the visibility between FP resonances across the main DP and the satellite DP within the inner cavity for $n_{\text{in}}>0$ and $n_{\text{in}}<0$ separately ($\mu_{e}$\SI{\sim 100000}{cm^2 V^{-1} s^{-1}}, $\mu_{h}$\SI{\sim 50000}{cm^2 V^{-1} s^{-1}}.
Charge carriers for $n_{\text{in}}<0$ (hole side) bouncing between boundaries formed by the satellite DPs ($\Delta G\sim0.9\%$) show a visibility that is 40$\%$ lower than the visibility of the main DP oscillation ($\Delta G \sim1.5\%$), as shown in Fig.~\ref{fig:Moiree_PNP}a. In contrast, for $n_{\text{in}}>0$ (electron side), the visibility is reduced by 85$\%$ ($\Delta G\sim0.4\%$ at the satellite DP compared to $\Delta G\sim3\%$ at the main DP).
To interpret these observations, we consider a family of possible Moir\'{e} minibands for graphene on hBN substrate, which was calculated by Wallbank \textit{et al.} \cite{Wallbank13} using a general symmetry-based approach. Because the relatively large number of model-dependent parameters of the symmetry breaking potential
can strongly influence the obtained Moir\'{e} perturbation, the focus was on the generic features for different sets of parameters used. In Fig.~\ref{fig:Moiree_PNP}b, the DOS for three different sets of parameters are plotted. Details on the parameters can be found in Ref. \cite{Wallbank13}.

For the case of $n_{\text{in}}<0$, when crossing the satellite DP on the hole side, the DOS between inner and outer cavity decreases to zero (independent of the parameters used for the calculations of the DOS as given in Ref. \cite{Wallbank13}) for all $k_{x,y}$, as indicated with the purple triangle in Fig.~\ref{fig:Moiree_PNP}b. This result is experimentally supported by capacitance spectroscopy of hBN-graphene-hBN heterostructures in the presence of a Moir\'{e} superlattice \cite{Yu14}. The vanishing DOS leads to similar reflection/transmission coefficients as for the main DP, resulting in a comparable visibility of the two FP resonances. A representative cut at fixed $n_{\text{out}}<0$ is shown in Fig.~\ref{fig:Moiree_PNP}d.

For the electron side with $n_{\text{in}}>0$, the significantly reduced visibility is in qualitative agreement with the band-structures shown in Fig.~\ref{fig:Moiree_PNP}b, since the DOS at the satellite DP (indicated with an orange triangle) is reduced (depending on the parameters used in the calculation), but never vanishes. A representative cut at fixed $n_{\text{out}}>0$ is shown in Fig.~\ref{fig:Moiree_PNP}c. A direct implication of the finite DOS is that only some charge-carriers have a nonzero reflection coefficient, thus contributing to the FP resonances, while the remaining ones account for a background current. It is worth noting, that even though the visibility of the FP resonances across the electron side satellite DP ($n_{\text{in,out}}>0$, Fig.~\ref{fig:Moiree_PNP}c) are significantly reduced, they seem to be more regular over a wider gate-range compared to the FP resonances across the hole satellite DP ($n_{\text{in,out}}<0$, Fig.~\ref{fig:Moiree_PNP}d). However, this could be as well just sample-specific since the mobility in the hole side was significantly lower compared to the electron side.

We now compare the evolution of the position of the individual FP resonance peaks ($|n_{\text{out}}|\sim \text{const.}$) and their relative spacing $\Delta n_{\text{in}}$ as a function of $n_{\text{in}}$, which are plotted in the lower and upper panel of Fig.~\ref{fig:Moiree_PNP}e. The latter analysis is performed instead of the cavity length analysis since Equation~\ref{cavity-length} does not hold any more if the Fermi-energy is tuned beyond the satellite DP. This is because in Equation~\ref{cavity-length} a circular Fermi-surface and electron like dispersion is assumed, which does not hold in the presence of a strong band-modulation. In contrast, mapping the FP resonance peak position as a function of charge-carrier density is free of any assumptions and therefore independent of the band-structure. If the semitransparent interfaces are defined via the main DP, the evolution of $\Delta n_{\text{in}}$ between neighboring peaks as a function of $n_{\text{in}}$ is in good agreement with the values extracted from theory (transport simulation in the absence of a Moir\'{e} superlattice), which is indicated with the black, dashed line in Fig.~\ref{fig:Moiree_PNP}e. If the semitransparent interfaces are defined via the satellite DPs, the density-spacing between the FP resonance peaks is further increased on both, the electron and hole side.

Although a precise extraction of the cavity size is not possible with Equation~\ref{cavity-length}, an estimate can still be given. We have found a cavity size of $L$\SIrange{\sim 250}{310}{nm} on the electron side if the density is measured from the main DP, whereas setting the density to zero at the satellite DP gives an unphysical cavity size ($L$\SIrange{\sim 80}{200}{nm}).  On the electron side the density of states seems almost unaltered, as can be seen in Fig.~\ref{fig:Moiree_PNP}b, however the band-structure is substantially modified, which can be seen in Ref.~\cite{Wallbank13}. The band-structure consists of two nonisotropic bands above the satellite DPs which makes the situation rather complex, with different visibilities and angle dependent transmissions for the two bands. 
On the hole side the cavity size analysis yields cavity sizes of $L$\SIrange{\sim 280}{360}{nm} if the density is measured from the main DP. At the satellite DP the density of states decreases to zero, meaning that a real Dirac point is formed. Therefore one might expect, that the density for the FP oscillations should be measured from the satellite DP. However by setting the charge-carrier density to zero at the hole satellite DP, the analysis gives unphysical results. We note, that close to the hole satellite DP Equation~\ref{cavity-length} should be valid if counting the charge-carrier density from the satellite DP, since the band-structure is isotropic. Besides the most pronounced FP resonances indicated with the purple arrow in Fig.~\ref{fig:Moiree_PNP}d, a second set of resonances with a much shorter period seem to appear in the very vicinity of the hole satellite DP (see inset of Fig.~\ref{fig:Moiree_PNP}d). For these resonances the extracted cavity length leads only to reasonable values ($L$\SI{\sim 300}{nm}) when counting the charge carrier density starting from the satellite DP. Unfortunately the residual doping ($n_{\mathrm{res}}$\SIrange{\sim 0.4e15}{1e15}{m^{-2}}) prevents us from resolving more of these features in the very vicinity of the satellite DP. A possible explanation for the different behavior of the two sets of FP oscillation ($n_{\text{out}}<0$) might be the following:
The small resonances are only observed up to densities of $\Delta n$\SI{<2e15}{m^{-2}} (where $\Delta n$ is measured from the satellite DP), where the Moir\'{e} miniband remains close to Dirac like (linear dispersion relation). The Fermi energy corresponding to theses small oscillations is consequently exclusively tuned in the linear part of the Moir\'{e} miniband (see Supporting Information).
From the band-structures we would expect that the small oscillations should be visible up to higher doping values, namely values which are larger by a factor of up to $\sim 10$ depending on the model. This estimate is based on the energy-spacing between the satellite DP and the first van Hove singularity at higher energies, extracted from theory (see Supporting Information). Note that the extracted energy-spacing corresponds to an upper bound within which the dispersion relation might be considered as linear. The Fermi energy corresponding to the stronger resonances indicated with the purple arrow, does partially reside outside of the linear region of the Dirac cones where the band-structure becomes more complex, including singularities and band-overlaps.

We note that in the work of Lee \textit{et al.} \cite{Lee16} they suggest that model 3, which is shown in Fig.~\ref{fig:Moiree_PNP}b, is the one describing the experimental situation best. Furthermore, for model 1, where multiple satellite DPs appear near every main DP, Equation~\ref{cavity-length} would also have to be multiplied by a factor of $\sqrt{g_{\text{SDP}}}$, where $g_{\text{SDP}}$ is the additional degeneracy of the satellite Dirac cones. Although by including additional degeneracies, and counting the density from the satellite DP, the strong oscillations could give reasonable cavity sizes, however, for the small oscillations, the additional degeneracies would result in too large cavity sizes. Moreover, most experimental evidence points to a single Dirac cone at the satellite DP.
We also note here, that the visibility of these resonances are quite low, which could also introduce some experimental error in the analysis.


In conclusion, we first analysed a \textit{pnp}-junction in the absence of a Moir\'{e} superlattice where the varying visibility of the different types of FP resonances could be linked to sharp and smooth \textit{pn}-junctions in the system. Furthermore, the change of the effective cavity length depending on $n_{\text{in}}$ and $n_{\text{out}}$ could be mapped via the FP resonances.
In a second sample, the presence of a Moir\'{e} superlattice gave rise to satellite DPs and the modified band-structure resulted in confinement of electronic trajectories and in the appearance of FP resonances. Although the oscillations are formed in the Moir\'{e} minibands, they can mostly be described as if they would originate from the non-modified band-structures. Further studies will be needed to explain these findings.

Our results show that confinement of electrons can be obtained using miniband engineering. Future studies can investigate the angle dependent transmission properties of such an interface. Moreover combining the studies of transverse magnetic focusing \cite{Lee16} and bias spectroscopy studies of FP resonances or detailed studies of the magnetic field dependence of FP resonances can reveal further details on the band reconstruction.\\

\textbf{ACKNOWLEDGMENTS}\\
This work was funded by the Swiss National Science Foundation, the Swiss Nanoscience Institute, the Swiss NCCR QSIT, the ERC Advanced Investigator Grant QUEST, and the EU flagship project graphene. M.-H.L. and K.R.  acknowledge  financial  support  by  the  Deutsche Forschungsgemeinschaft (SFB 689). The  authors  thank  John Wallbank, David Goldhaber-Gordon, Simon Zihlmann and L\'{a}szlo Oroszl\'{a}ny  for  fruitful discussions.\\


\bibliographystyle{PRL}


\clearpage
\onecolumngrid
        \setcounter{figure}{0}
        \renewcommand{\thefigure}{S\arabic{figure}}%
\section{Supporting information}

\subsection{Low magnetic field measurements}
\begin{figure}[htbp]
    \centering
      \includegraphics[width=1\textwidth]{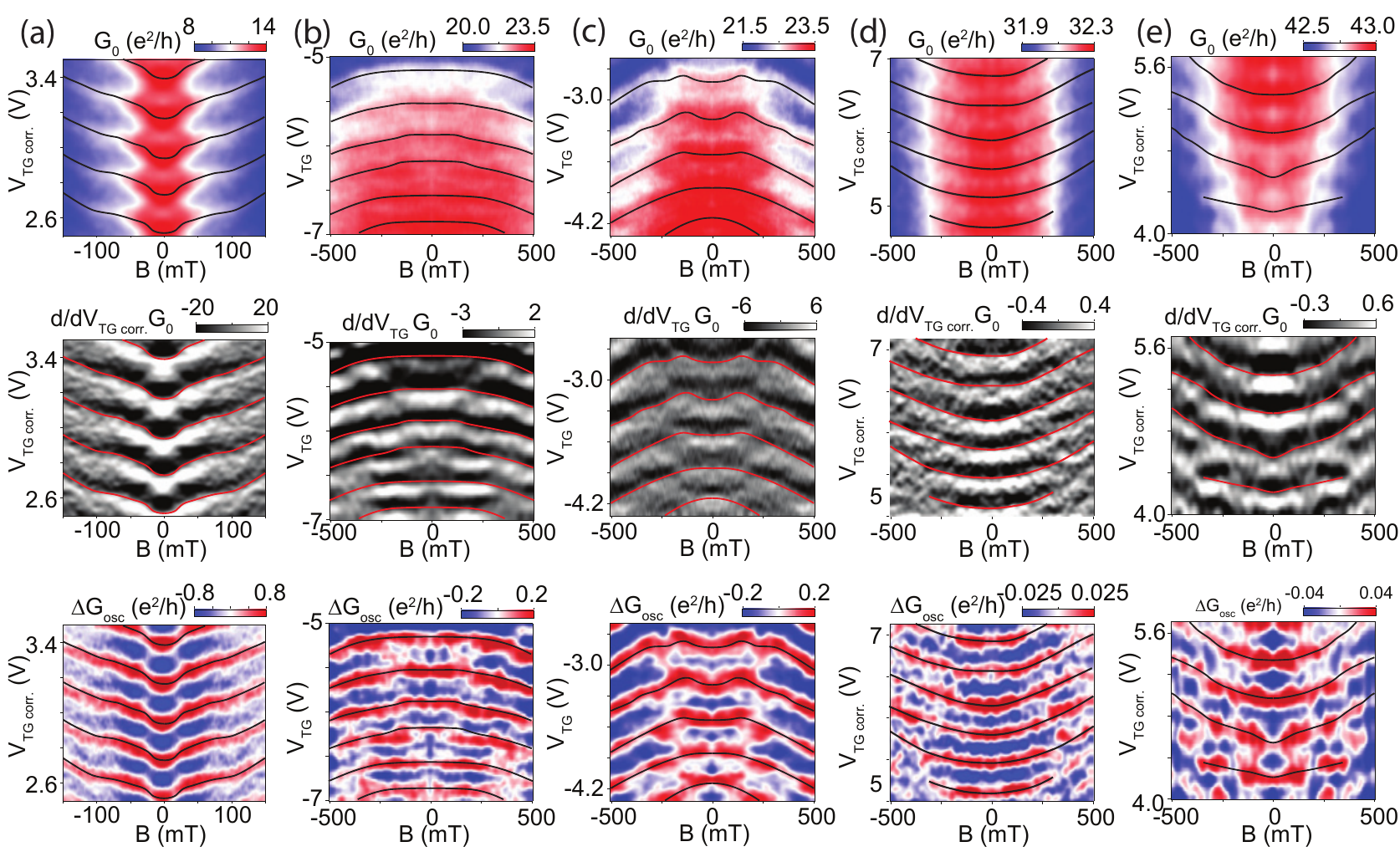}
    \caption{\textbf{Low magnetic field measurements of the FP resonances across the main and satellite DPs. First row} is the net conductance, \textbf{second row} the numerical derivative along the gate-axis and \textbf{third row} the net conductance oscillation. Columns \textbf{(a)} is the measurements across the main DP with $n_{\text{in}}>0$ and $n_{\text{out}}\sim-0.3\cdot10^{16}\,$m$^{-2}$, column \textbf{(b,c)} are measurements across the satellite DP with $n_{\text{in}}<0$ where $n_{\text{out}}\sim-0.7\cdot10^{16}\,$m$^{-2}$ and $n_{\text{out}}\sim-2.8\cdot10^{16}\,$m$^{-2}$, column \textbf{(d,e)} are measurements across the satellite DP with $n_{\text{in}}>0$ where $n_{\text{out}}\sim2.1\cdot10^{16}\,$m$^{-2}$ and $n_{\text{out}}\sim2.8\cdot10^{16}\,$m$^{-2}$.}
    \label{fig:Pi_shift}
\end{figure}


Fabry-Perot (FP) resonances disperse towards higher doping ($k$ values) with increasing magnetic field \cite{Young09,Shytov08,Rickhaus15,Rickhaus13,Grushina13,Calado15} as seen in Fig.~\ref{fig:Pi_shift}a-c. This results from the resonance condition $\Delta \theta_{\text{WKB}}+\Phi_{\text{AB}}= \mathrm{const}.$  where $\Delta \theta_{\text{WKB}}$ is the Wentzel-Kramers-Brillouin phase acquired on the path of the charge-carriers trajectory, and $\Phi_{\text{AB}}$ is the Aharonov Bohm phase \cite{Rickhaus15, Varlet16}.

Furthermore one can determine the Berry's phase acquired by a charge carrier encircling the origin in k-space. For single layer graphene with a staggered sublattice potential, which breaks its inversion symmetry \cite{Zhou07}, the Berry's phase becomes exactly $\pi$ as the gap approaches zero \cite{Xiao10}. This has been observed experimentally in FP measurements \cite{Young09} and quantum Hall measurements \cite{Novoselov05}. In Fig.~\ref{fig:Pi_shift} we show low magnetic field measurements for FP oscillations across the main DP (a) and the satellite DP in the hole (b,c) and electron (d,e) doped region. Each of the measurements was plotted in three different ways showing the net conductance, the numerical derivative of the net condctance and the net oscillation. The black/red lines in Fig.~\ref{fig:Pi_shift} were extracted numerically and track the peak-position of the FP oscillations with magnetic field. For the FP resonances across the main DP we see the $\pi$-shift at roughly $B=30-50\,$mT as shown in Fig.~\ref{fig:Pi_shift}a. Following the previously given argument, one  might expected as well a $\pi$-shift for the FP resonances across the satellite DP on the hole side, since there the DOS drops to zero comparable to the main DP. However, measurments at different $n_{\text{out}}$, shown in Fig.~\ref{fig:Pi_shift}b,c, did not allow a conclusive statement whether such a $\pi$-shift is present or not. It seems however that for both of the here preseneted situations an intriguing pattern appears at low magnetic fields (i.e. $B<200\;$mT). At higher field, the FP oscillations disperse constantly. On the electron side, i.e. $n_{\text{in}}>0$, the situation is less clear. The pattern in the middle exhibits less strong variations.

\subsection{Bias-spectroscopy of Fabry-P\'{e}rot resonances}
\begin{figure}[htbp]
    \centering
      \includegraphics[width=1\textwidth]{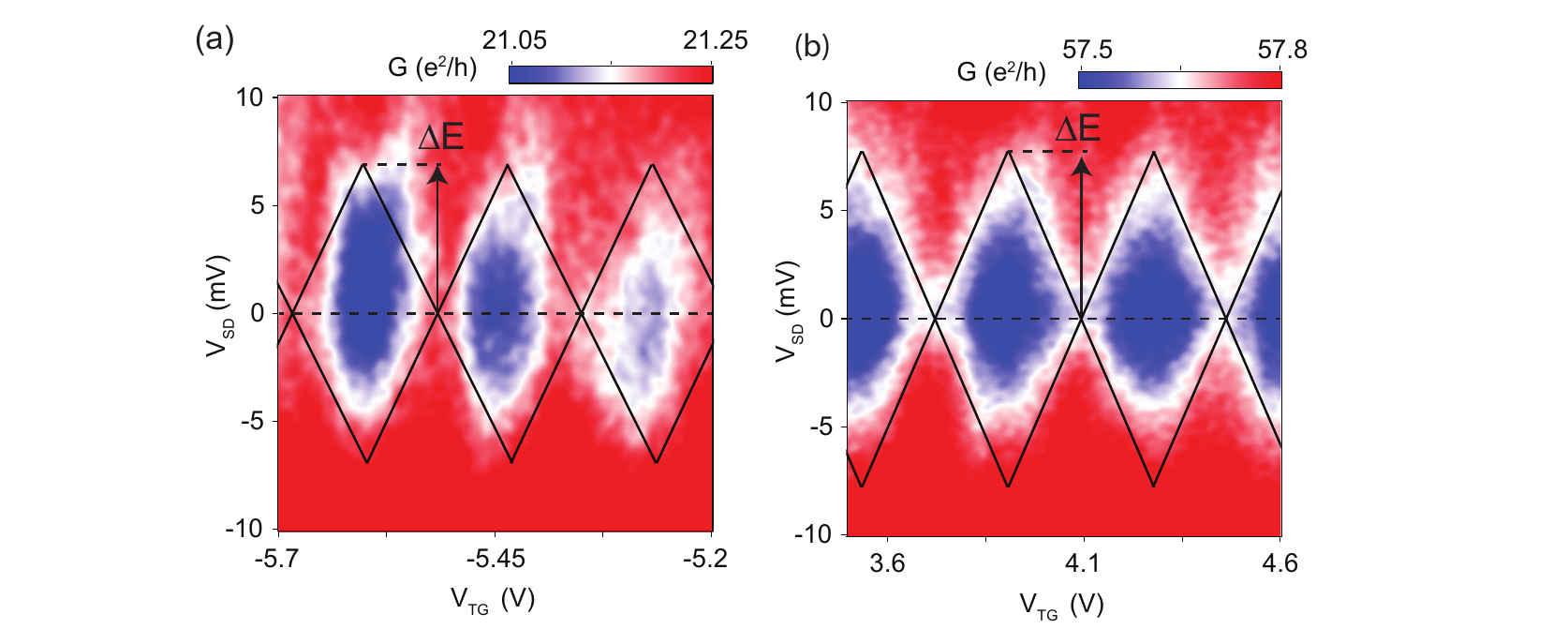}
    \caption{\textbf{Bias spectroscopy on FP resonances a,} Measurement of FP resonances across the main DP ($n_{\text{in}}<0$, $n_{\text{out}}>0$) where an energy spacing corresponding to $L=300\,$nm is indicated. \textbf{b,} Comparable measurement for the satellite DP with $n_{\text{in}},n_{\text{out}}>0$. Energy spacing corresponds to $L=280\,$nm.}
    \label{fig:SI_Vsd_meas}
\end{figure}
Bias spectroscopy of the FP resonances can be used as an alternative way to extract the cavity length. The resonance condition in a cavity is given by $L=j \cdot \lambda/2$, or $k_{\text{F}}=j \pi/L$, where $j$ is an integer. For single-layer graphene in the absence of a Moir\'{e} superlattice, the energy spacing $\Delta E$ between two consecutive FP resonances is equidistant and given by
\begin{equation}
\Delta E=\hbar v_{\text{F}} [(j+1)-j] \frac{\pi}{L}=\hbar v_{\text{F}} \frac{\pi}{L}
\end{equation}
where $\hbar$ is the reduced Planck constant and $v_{\text{F}}\sim10^6\,$m/s is the Fermi velocity. The size of the diamond corresponds to twice the level spacing, $2 \Delta E$, leading to:
\begin{equation}\label{equation_energy}
L=\hbar v_{\text{F}} \frac{\pi}{\Delta E}.
\end{equation}

The measurement of the bias spectroscopy across the main DP is in qualitative agreement with the extracted cavity length using Equation~(\ref{Cavity_length}). In  Fig.~\ref{fig:SI_Vsd_meas}a, an energy spacing corresponding to $L\sim300\,$nm according to Equation~(\ref{equation_energy}) is indicated. Similar results are found for the FP resonances across the satellite DP with $n_{\text{in}}>0$, shown in Fig.~\ref{fig:SI_Vsd_meas}b where an energy spacing corresponding to  $L=280\,$nm is indicated in the plot. Note that Equation~(\ref{equation_energy}) is based on the energy-dispersion of unperturbed graphene, neglecting a modification of the band structure in the presence of a Moir\'{e} superlattice. On the hole side, the pattern was too complex to extract any reasonable information.
\newpage

\subsection{Quantum Hall Measurements}
\begin{figure}[htbp]
    \centering
      \includegraphics[width=1\textwidth]{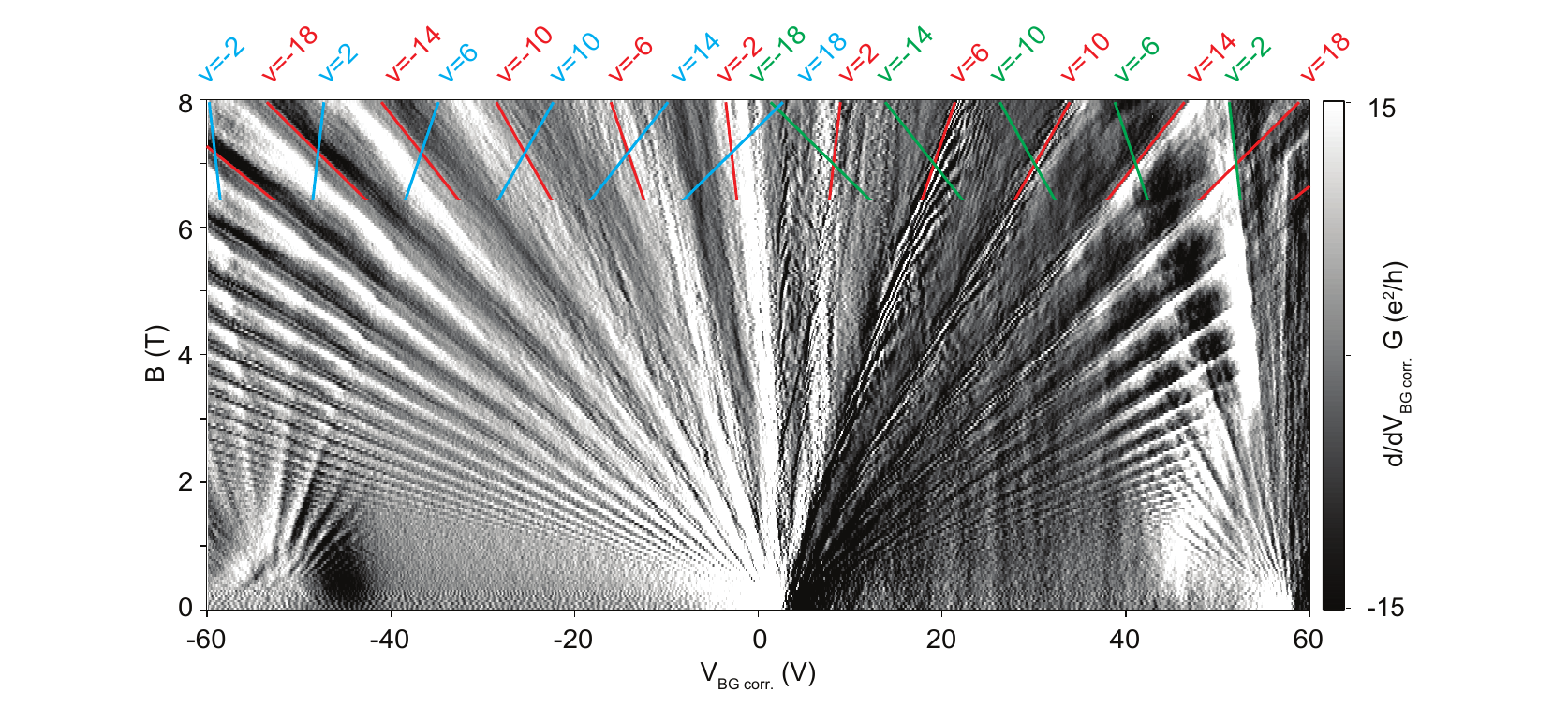}
    \caption{\textbf{Quantum Hall measurement in high magnetic field.} Numerical derivative of the conductance as a function of the global back-gate and magnetic field reveals the expected filling-factors emerging at the main DP (red) and at the satellite DPs (cyan and green) as observed in Ref.~\cite{Dean13,Ponomarenko13}}
    \label{fig:QHE}
\end{figure}
Measurement of the quantum Hall Effect in the unipolar regime reveals the emergence of the filling-factors associated with graphene from the main DP and the two satellite DPs, and similar results have been reported in Ref.~\cite{Dean13,Ponomarenko13}.

\subsection{Thermal annealing of hBN-graphene-hBN heterostructures}
\begin{figure}[htbp]
    \centering
      \includegraphics[width=1\textwidth]{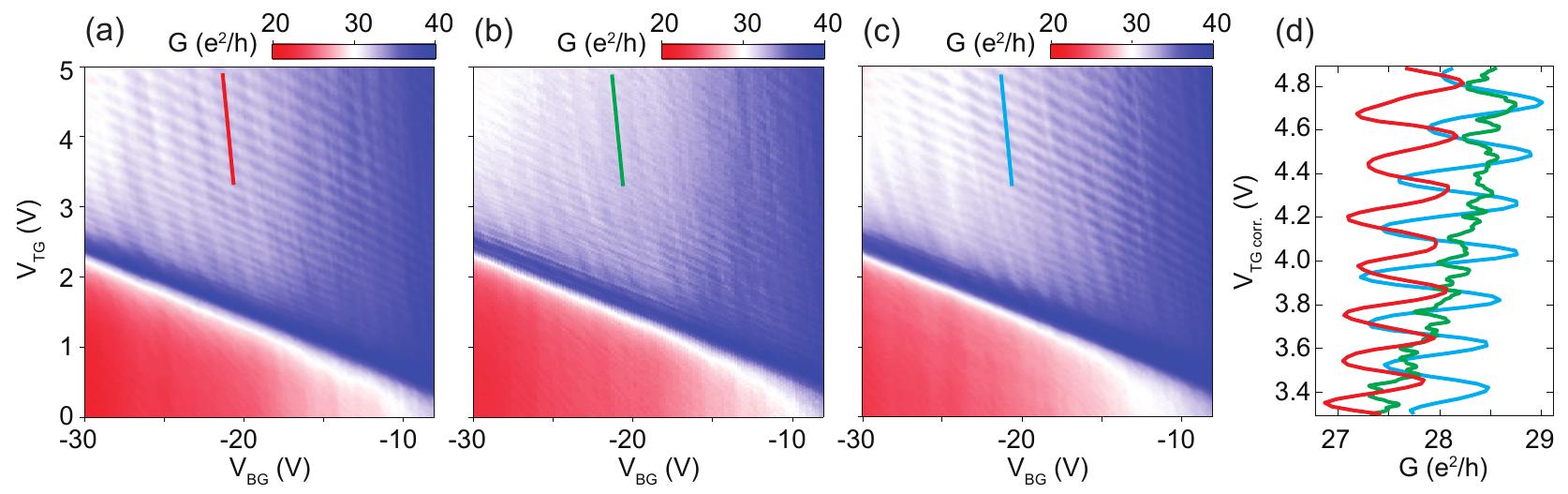}
    \caption{\textbf{Self-cleaning properties of hBN-graphene-hBN heterostructures upon thermal annealing. a,} Conduction map with FP resonances clearly visible in the bipolar region. \textbf{b,} Same map after degradation of the sample where the FP resonances are nearly absent. \textbf{c,} Most of the FP resonances are restored after thermal annealing at 200 $^\circ$C for 20 min (outside the cryostat). \textbf{d,} Linecuts as indicated in (a)-(c) in comparison.}
    \label{fig:Self-healing}
\end{figure}
As a result of applying high gate voltages the device quality degraded over a longer time period (2 weeks). The degradation over time and the subsequent improvement upon thermal annealing was observed in several samples independently. In between the maps shown in Fig.~\ref{fig:Self-healing}a and \ref{fig:Self-healing}b, the global and local gates were swept over a wide range (e.g. $\pm60\,$V for the global back-gate during the QHE measurement) for an extended time period (2 days). The sample shown in the main text was annealed at least 6 times outside the cryostat. No significant difference between annealing on a hotplate in air at $200-250\,^{\circ}$C or in a rapid thermal annealer under forming-gas atmosphere (Ar/H$_2$) at $300\,^{\circ}$C  could be found. An example before and after thermal annealing is shown in Fig.~\ref{fig:Self-healing}a-c. In Fig.~\ref{fig:Self-healing}d an identical linecut of the original (red), degraded (green) and annealed (blue) sample are shown in comparison. We speculate that the decrease of the device quality might be due to the migration of contaminations when continuously sweeping the gates to high voltages. By applying a thermal annealing step, these contaminations are very likely to aggregate again in pockets resulting in larger areas of clean graphene \cite{Kretinin14,Engels14}.

\subsection{Temperature dependence of Fabry-P\'{e}rot resonances}
\begin{figure}[htbp]
    \centering
      \includegraphics[width=1\textwidth]{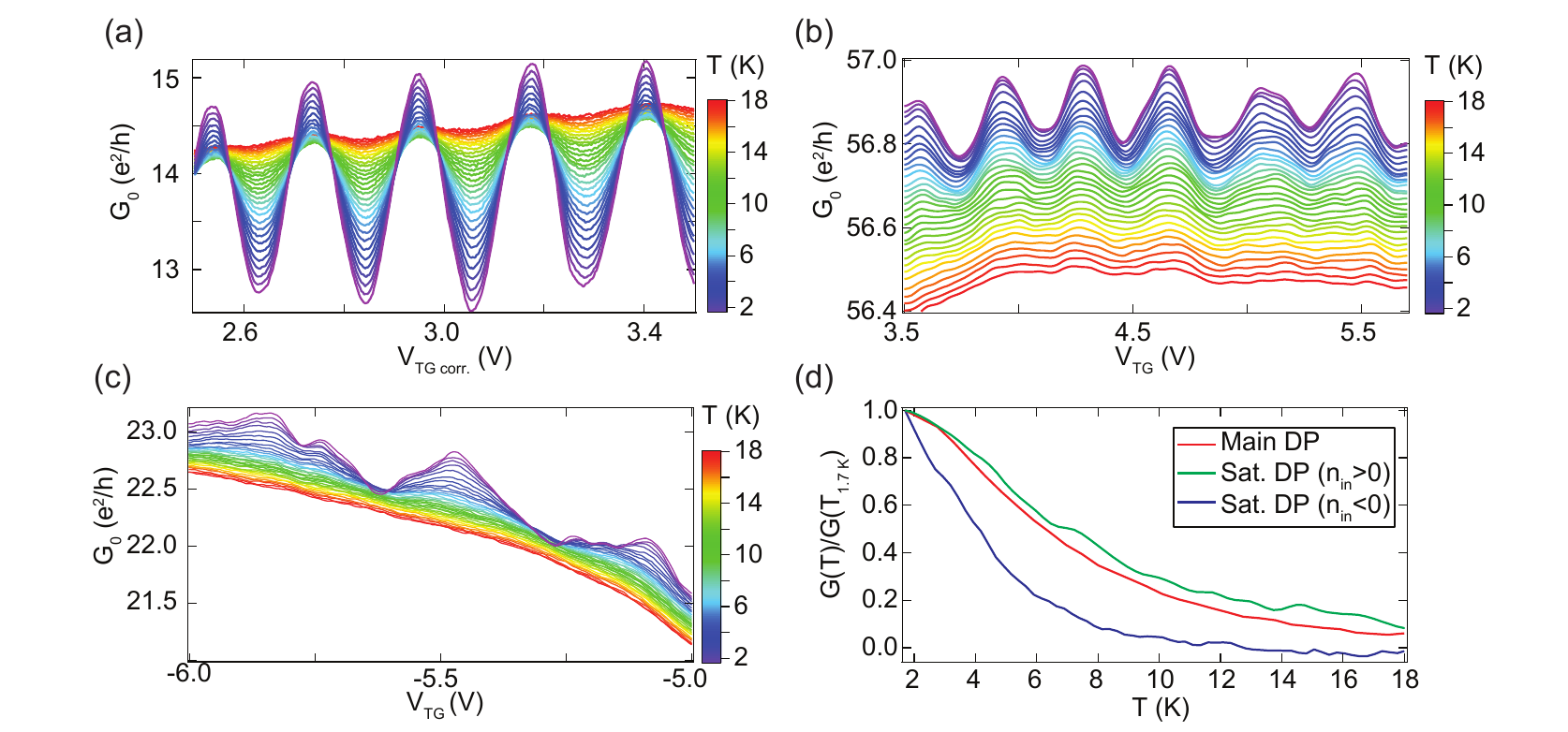}
    \caption{\textbf{Temperature dependence of the FP resonances} FP resonances across the \textbf{a,} main DP and the two satellite DPs where \textbf{b,} $n_{\text{in}}<0$ and \textbf{c,} $n_{\text{in}}>0$ as a function of temperature. \textbf{d,} Oscillation amplitude as a function of temperature renormalized by the value measured at base temperature ($T\sim1.7\,$K). While the FP resonances remains very fainth even at $T\sim20\,$K for (a) and (b), it disappears already around $T\sim10\,$K in the case of (c).}
    \label{fig:FP_Moiree_Temp}
\end{figure}
The temperature dependence of the various FP resonances are shown in Fig.~\ref{fig:FP_Moiree_Temp}. Across the main DP and the satellite DP on the electron side (Fig.~\ref{fig:FP_Moiree_Temp}a,b), the temperature where the resonances disappear are in the order of $T\sim20\,$K ($E\sim2\,$meV), which is on the same order of the level spacing obtained from bias spectroscopy. The FP resonances across the satellite DP on the hole side seem to vanish at slightly lower temperatures ($T\sim10\,$K, $E\sim1\,$meV).
\newpage

\subsection{Fabrication}
\begin{figure}[htbp]
    \centering
      \includegraphics[width=1\textwidth]{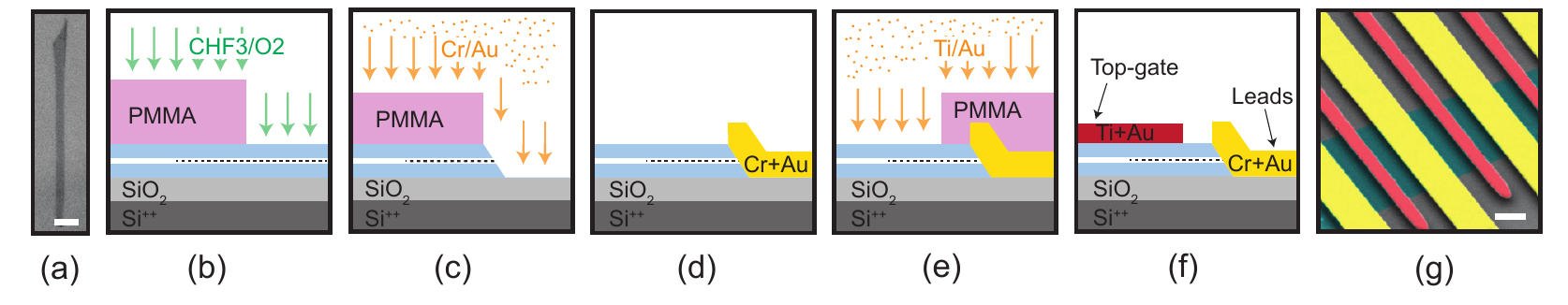}
    \caption{\textbf{Fabrication of a 2-terminal \textit{pnp}-junction array. a,} Exfoliated graphene flake having a width of $\sim1.5\,\mu$m. Scale-bar equals $5\,\mu$m. \textbf{b-d,} The assembly of the hBN-graphene-hBN heterostructure and subsequent etching of the side-contacts follows mostly the procedure described in \cite{Wang13}. \textbf{e,f,} Establishing the local top-gates without the need to etch the encapsulated graphene. \textbf{g,} False-color SEM image with side-contacts (yellow), top-gates (red) and the graphene encapsulated between two layers of hBN (cyan). Scale-bar equals $500\,$nm.}
    \label{fig:Fabricatioin}
\end{figure}
The fabrication of the hBN-graphene-hBN heterostructure follows Ref.~\cite{Wang13} in most steps with some variations and extensions as explained in the following. Exfoliation is done on a  Si$^{++}$/SiO$_2$ substrate with a 315$\,$nm-thick oxide, using the scotch-tape technique. The chips were previously cleaned using Piranha solution (98\% H$_2$SO$_4$ and 30\% H$_2$O$_2$ in a ratio of 3:1). For the assembly of the heterostructure, we choose only graphene-flakes having a width not exceeding 2 $\mu$m as shown in Fig.~\ref{fig:Fabricatioin}a. This is beneficial when fabricating the local top-gates. Next, self-aligned side-contacts are established to the graphene as shown in Fig.~\ref{fig:Fabricatioin}b-d. Self-aligned means that the same PMMA mask is used for etching the hBN-graphene-hBN heterostructure and subsequent evaporation of the Cr/Au ($10\,$nm/$50\,$nm) contacts. This leads to very transparent contacts ($\sim50-100\,$ $\Omega\cdot \mu$m) since the exposed graphene edge never comes into contact with any solvent or polymer. It is worth mentioning that we use cold-development ($T\sim3-5\,^{\circ}$C) with IPA:H$_2$O (7:3) to reduce cracking of the PMMA on hBN \cite{Lee16,Rooks02}. In the last step, the local top-gates  are evaporated (Fig.~\ref{fig:Fabricatioin}e). Since a narrow graphene flake was chosen in the very beginning of the fabrication, no etching step to shape the graphene is needed. The spacing between the top-gate and graphene is defined only by the top-hBN layer which can be chosen very thin. Therefore we can achieve high carrier concentrations and we are able to establish relativly sharp \textit{pn}-junctions.

In order to fabricate the Moir\'{e} superlattice, straight edges of the graphene and the top-hBN are aligned with respect to each other \cite{Geim07}. With this technique, a rotation angle between top-hBN and graphene of less than 1 degree can be achieved with a corresponding superlattice period of $\sim10\,$nm \cite{Yankowitz12}. The aligned graphene-hBN stack is then placed down a large angular misalignment.

\subsection{Theoretical Model}
\begin{figure}[htbp]
    \centering
      \includegraphics[width=1\textwidth]{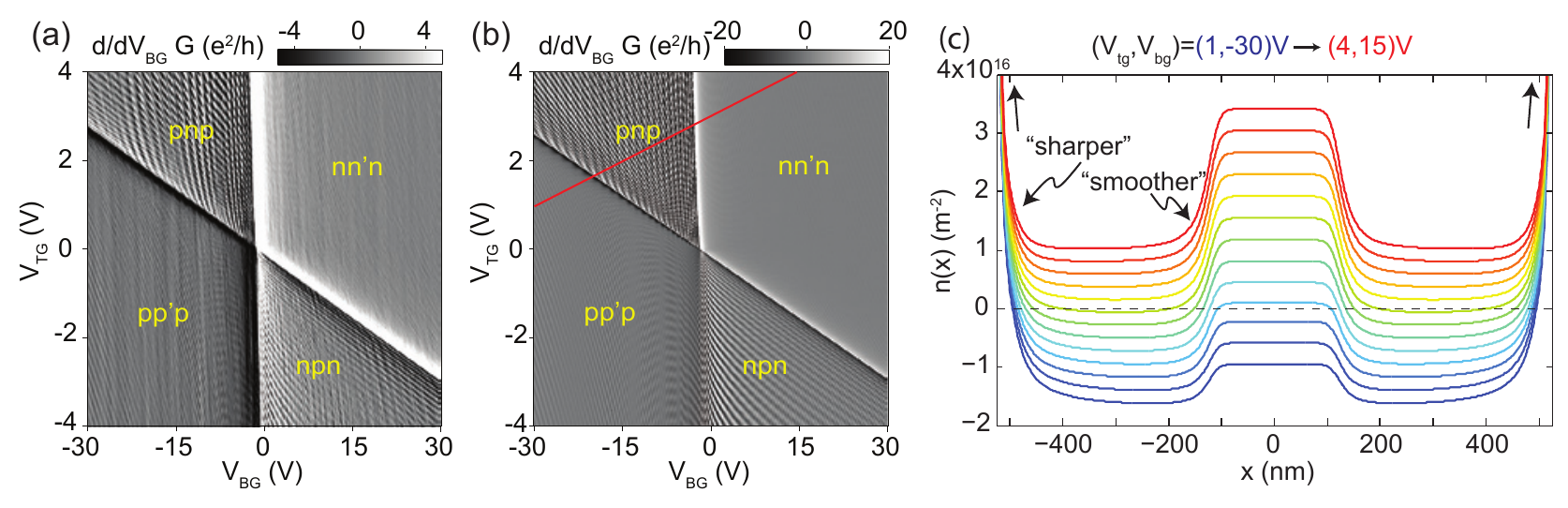}
    \caption{\textbf{Theoretical transport simulation. a,} The numerical derivative of the conductance as a function of global back-gate and local top-gate of the experimental data.  \textbf{b,} Comparable map as in (a) obtained from transport simulation which reproduces the experimental data very well. \textbf{c,} Calculated density-profiles throughout the sample for different gate-voltages as indicated in (b) with red.}
    \label{fig:SI_Exp_vs_Thory}
\end{figure}

The transport simulation used in this study is identical to the one used in Ref.~\cite{Rickhaus13}. More information on the calculations are summarized in the supplementary information of Ref.~\cite{Rickhaus13} or in more detail in Ref.~\cite{Liu12_2,Liu13_2}. We note that in order to have constructive/destructive interference, two interfaces with a non-zero reflection coefficient are required. Since the contacts dope the graphene in their vicinity into $n$ type, we observe stronger oscillations in the \textit{pnp} and \textit{ppp} compared to the \textit{npn} and \textit{nnn} cases. The better visibility is due to the additional \textit{pn}-junction close to the contact, which leads to larger reflectivity. This can be seen both in experiment and simulation.

\subsection{Sharp vs. smooth \textit{pn}-junction}
A symmetric \textit{pn}-junction is commonly regarded as sharp or smooth by the definition of $\lambda_{\text{F}}\gg d$ or $\lambda_{\text{F}}\ll d$, respectively, where $d$ is the transition region from \textit{n}- to \textit{p}-doping and $\lambda_{\text{F}}=2 \sqrt{\pi/n}$ is the Fermi wavelength far from the \textit{pn}-junction, where the carrier density is homogeneous.

\subsection{Definition of the cavity length}
\begin{figure}[htbp]
    \centering
      \includegraphics[width=1\textwidth]{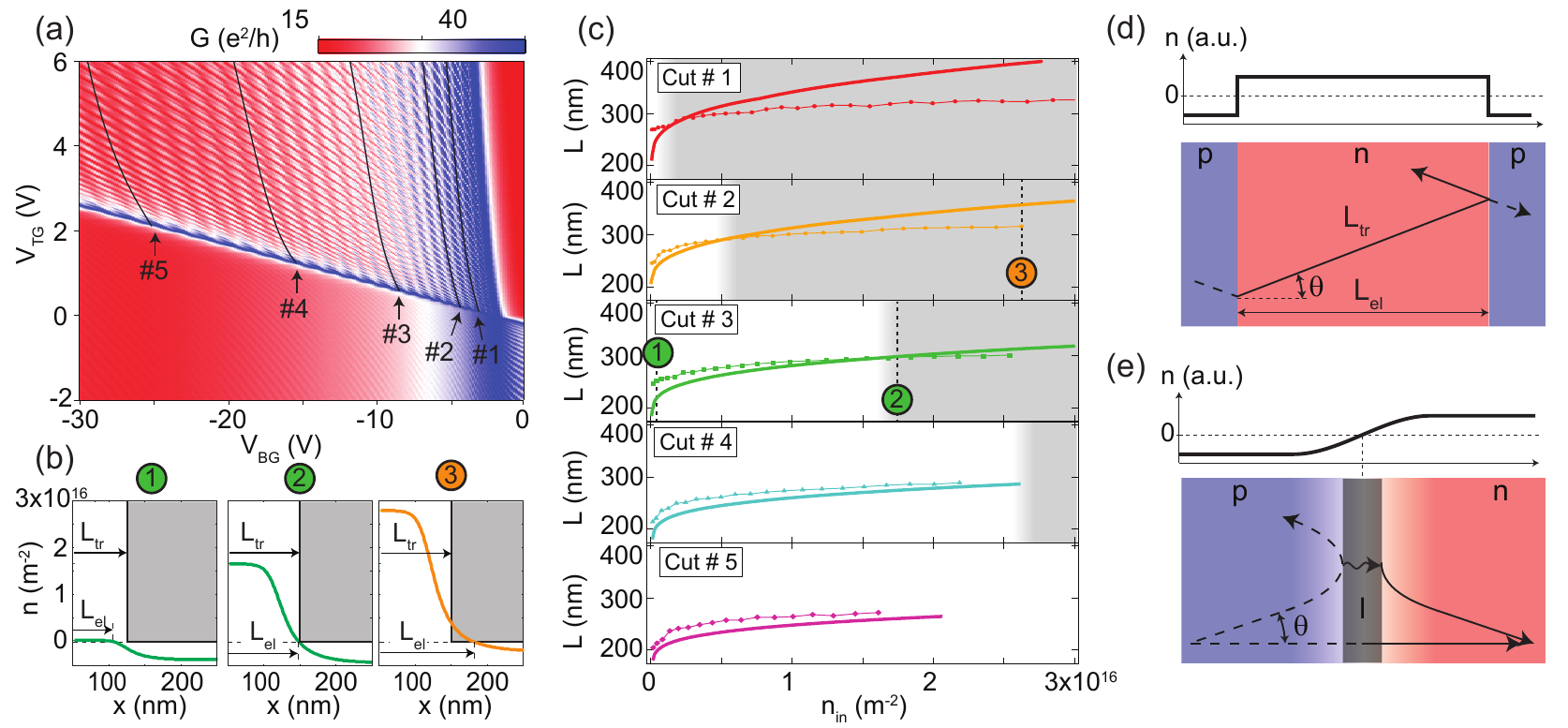}
    \caption{\textbf{Definition of the cavity length. a,} Transport simulation reproducing the Fabry-Perot pattern observed for the regular \textit{pnp}-junction in the experiment. The linecuts indicated with 1-5 are at identical positions with those extracted from the experiment. \textbf{b,} Three different density profiles where $L_{\text{tr}}>L_{\text{el}}$, $L_{\text{tr}}=L_{\text{el}}$ and $L_{\text{tr}}<L_{\text{el}}$. The inner cavity is centred around $x=0\,$nm. \textbf{c,} Comparison of the extracted cavity length from the transport simulation ($L_{\text{tr}}$) using Equation~\eqref{Cavity_length} (line + markers) and the cavity length extracted from electrostatic calculations ($L_{\text{el}}$, bold line) for different cuts as indicated in (a). Depending on the inner inner and outer cavity-doping, $L_{\text{tr}}$ can be larger or smaller than $L_{\text{el}}$. The two models are in best agreement for high doping of $n_{\text{in}}$ and $n_{\text{out}}$ such as shown for cut 5. \textbf{d,e,} Further small corrections which are responsible that the extracted cavity length $L_{\text{tr}}$ deviates from $L_{\text{el}}$.}
    \label{fig:Cavity_length_Theory}
\end{figure}

In the main text, the cavity length was extracted from consecutive peaks of constructive interference in the measurement. However, the extracted cavity length in this case does not correspond to the distance between the two points of zero charge-carrier density of the left and right \textit{pn}-interface, as one might think. Here we elaborate on the difference between the cavity length extracted from transport measurements and from electrostatic considerations. Furthermore, two additional aspects which account for minor corrections on the cavity length are discussed.

\subsection{Cavity length from electrostatic calculations ($L_{\text{el}}$)}
Probably the most straight-forward definition of the cavity length in a \textit{pnp}-junction is by the distance between the two points where the charge-carrier density is zero. In the simulation, for every set of ($V_{\text{BG}}$,$V_{\text{TG}}$) a density profile along the x-axis (defined perpendicular to the \textit{pn}-junction, $x=0$ is centred in the middle of the top-gate) was calculated based on the quantum capacitance model for graphene\cite{Liu13_2} with classical self-partial capacitances simulated using FEniCS\cite{FEniCS} and Gmsh\cite{Gmsh}.

Three exemplary profiles (zoom near the right \textit{pn}-junction of the inner cavity) are shown in Fig.~\ref{fig:Cavity_length_Theory}b and further examples can be seen in Fig.\ref{fig:SI_Exp_vs_Thory}c. Finally the evolution of $L_{\text{el}}$ (inner cavity) as a function of $n_{\text{in}}$ for the linecuts indicated in Fig.~\ref{fig:Cavity_length_Theory}a is shown in Fig.~\ref{fig:Cavity_length_Theory}c (bold lines). The transport simulation in Fig.~\ref{fig:Cavity_length_Theory}a is based on the same electrostatic model.

\subsection{Cavity length from transport measurements ($L_{\text{tr}}$)}
\begin{figure}[htbp]
    \centering
      \includegraphics[width=1\textwidth]{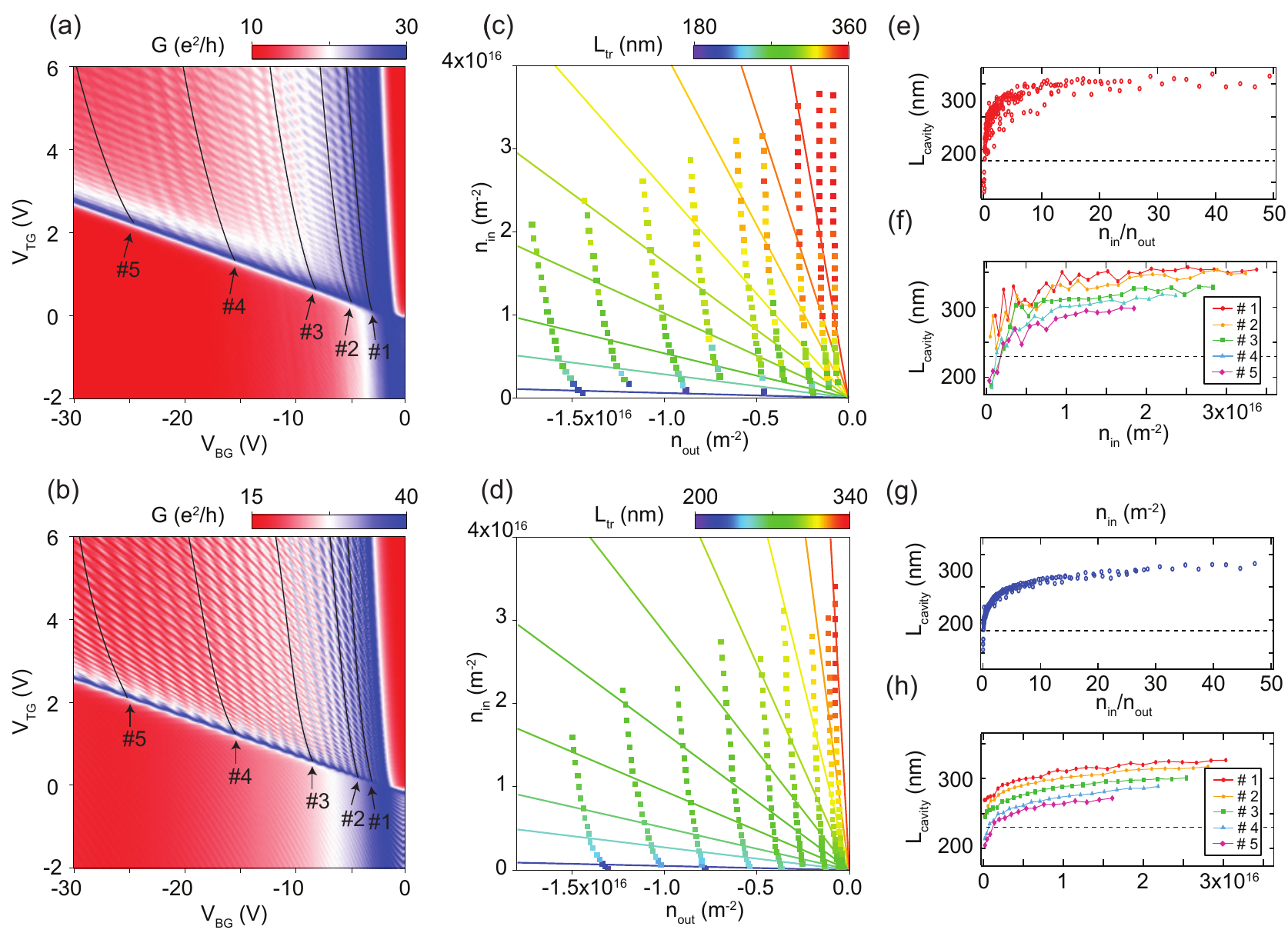}
    \caption{\textbf{Cavity length as a function of ($\mathbf{n_{\text{in}}}$,$\mathbf{n_{\text{out}}}$) or $\mathbf{n_{\text{in}}/n_{\text{out}}}$ for experiment (top row) and theory (bottom row). a,b,} Conduction map as a function of $V_{\text{BG}}$ and $V_{\text{TG}}$. Selected linecuts are indicated. \textbf{c,d,} $L_{\text{tr}}$ as a function of $n_{\text{in}}$ and $n_{\text{out}}$. The solid lines are a guide to the eye for a constant cavity length. \textbf{e,g,} The same data as in (c,d) plotted as the ratio $n_{\text{in}}/n_{\text{out}}$. \textbf{f,h,} Selected linecuts as indicated in (a,b) at fixed $n_{\text{out}}$.}
    \label{fig:Matrix_CavL}
\end{figure}
The cavity length extracted from the position of two neighbouring FP peaks, assuming a FP resonator with a hard-wall potential, is given by
\begin{equation}\label{Cavity_length}
L_{\text{tr}}=\frac{\sqrt{\pi}}{\sqrt{n_{\text{j+1}}}-\sqrt{n_{\text{j}}}}
\end{equation}
as derived in the main text. Using a box-shaped potential is an oversimplification and can lead to a difference compared to $L_{\text{el}}$. In Fig.~\ref{fig:Cavity_length_Theory}b the calculated $L_{\text{tr}}$ is sketched in direct comparison with the calculated density-profile from electrostatics. In Fig.~\ref{fig:Cavity_length_Theory}c $L_{\text{tr}}$ (line + symbols) is plotted together with $L_{\text{el}}$ (bold lines) where the shaded region account for $L_{\text{el}}>L_{\text{tr}}$.

The cavity length $L_{\text{tr}}$ is fixed for a given ratio of $n_{\text{in}}/n_{\text{out}}$ as explained in the main text. The same is true for $L_{\text{el}}$, because if $n_{\text{in}}$ and $n_{\text{out}}$ are varied by a common factor, only the slope of the \textit{pn}-junction changes, whereas the position of zero density remains constant. In Fig.~\ref{fig:Matrix_CavL} $L_{\text{tr}}$ is plotted (for experiment and theory) in various ways to see the different trends more clearly. In Fig.~\ref{fig:Matrix_CavL}a/b the conduction map as a function of the global back-gate and local top-gate is given with selected linecuts indicated with $\#1-\#5$. $L_{\text{tr}}$ extracted from various linecuts is plotted in Fig.~\ref{fig:Matrix_CavL}c/d as a function of $n_{\text{in}}$ and $n_{\text{out}}$, the positions of the corresponding FP resonances. The solid lines are a guide to the eye where $L_{\text{tr}}$ is constant. It can be easily seen that the condition $L_{\text{tr}}(n_{\text{in}}/n_{\text{out}})=\text{const.}$ is well fulfilled since the colors of the points showing $L_{\text{tr}}$ remain constant along the solid lines. The same information is replotted in Fig.~\ref{fig:Matrix_CavL}e/g, where all points of Fig.~\ref{fig:Matrix_CavL}c/d are plotted as a function of the ratio $n_{\text{in}}/n_{\text{out}}$. All values $L_{\text{tr}}$ follow the same curve, independent of the position ($n_{\text{in}},n_{\text{out}}$). In Fig.~\ref{fig:Matrix_CavL}f/h selected linecuts ($n_{\text{out}}=\text{const.}$, as indicated in Fig.~\ref{fig:Matrix_CavL}a/b) are plotted against $n_{\text{in}}$. From these curves the trends of $L_{\text{tr}}$ for fixed (i) $n_{\text{out}}$ or (ii) $n_{\text{in}}$ was deduced as described in the main text.

\subsubsection{Second order corrections}
Besides the major difference of assuming a box-shaped hard-wall potential, there are further smaller corrections leading to a difference between $L_{\text{tr}}$ and $L_{\text{el}}$:
\begin{itemize}
\item Assume there is a density-profile as indicated in Fig.~\ref{fig:Cavity_length_Theory}d. Since the trajectories contributing most to the FP signal have a finite incident angle $\theta$ (for $\theta=0$ one ends up with Klein-tunneling, thus no contribution to the FP \cite{Kartnelson06,Robinson12,Klein29,Stander09}), the extracted $L_{\text{tr}}$ actually corresponds to the diagonal distance. Therefore the real cavity length (which is given in this case by $L_{\text{el}}$) is given by $L_{\text{el}}=L_{\text{tr}} \cos(\theta)$. In this case $L_{\text{tr}}$ over-estimates the real cavity size. Since the charge-carriers with a small incident angle (with respect to the \textit{pn}-junction normal) account for most of the FP signal, this results only in a minor correction.
\item As already mentioned before, the density profile is smooth and not abrupt. This leads to a bending of the charge-carrier near the $n=0$ density line as sketched in Fig.~\ref{fig:Cavity_length_Theory}e. Charge carrier are therefore reflected before they hit the $n=0$ density line, which will make the effective cavity size shorter. The distance $l$ can be calculated according to $l=v p_{\text{y}}/E_{\text{x}}$ \cite{Cheianov06}. However, since the trajectories of the charge carriers are bent, and in addition the density is varying while approaching the \textit{pn}-junction, it is hard to make any statements if this effect will lead to an over- or under-estimation of the cavity size.
\end{itemize}

\subsection{Conclusion}
Using Equation~(\ref{Cavity_length}) to extract the cavity length from FP resonances, gives slightly different results than using the definition based on simulated carrier densities. The extracted cavity lengths $L_{\text{tr}}$ and $L_{\text{el}}$ are only in good agreement (over a longer density range) when the outer and inner cavities are highly doped. In our experiment this condition is best satisfied for Cut 5 as shown in Fig.~\ref{fig:Cavity_length_Theory}c. This can be understood since at high doping the transition from \textit{p}- to \textit{n}- region is sharper than for low doping, thus best resembling a box-potential. Nevertheless, using Equation~(\ref{Cavity_length}) for the whole gate/density range is justified by the fact that for both, experiment and theory, the same quantity is extracted.

\subsection{Extracting the peak-position from Fabry-Perot resonances}
\begin{figure}[htbp]
    \centering
      \includegraphics[width=1\textwidth]{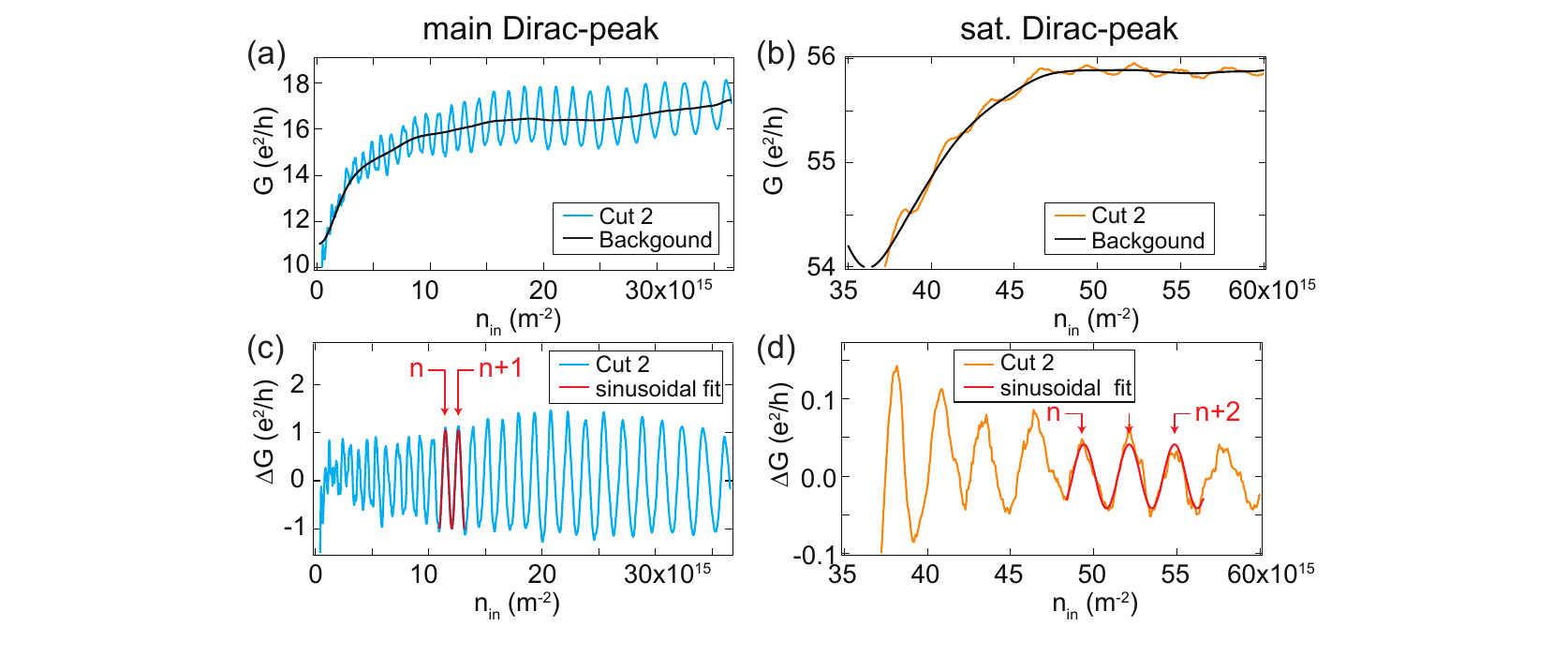}
    \caption{\textbf{Algorithm to extract the exact peak-position of the Fabry-Perot resonances of a \textit{pnp}-junction. a,} FP resonances (Cut 2, Fig.~\ref{fig:Matrix_CavL}a) across the main DP (cyan) and the corresponding background (black) obtained by smoothening.  \textbf{b,} FP resonances across the satellite DP (orange) where the inner cavity is tuned above the satellite DP with the corresponding background (black). \textbf{c,d,} Net oscillations ($\Delta G$) with sinusoidal fit to extract the exact peak-positions.}
    \label{fig:Extract_Peak_Pos}
\end{figure}
In order to extract the exact position of the FP resonances, the background of the line-profiles was subtracted by smoothening the linecut sufficiently (black line in Fig.~\ref{fig:Extract_Peak_Pos}a,b). The net resonances ($\Delta G$) were fitted with a sinusoidal curve depicted in Fig.~\ref{fig:Extract_Peak_Pos}c,d respectively. Since the FP oscillation of the satellite DP is much weaker and reveals a higher signal-to-noise ratio (compared to the FP measured across the main DP) the fitting-range was enlarged.

\subsection{Fabry-P\'{e}rot in inner and outer cavity across the satellite Dirac Peak}
\begin{figure}[htbp]
    \centering
      \includegraphics[width=1\textwidth]{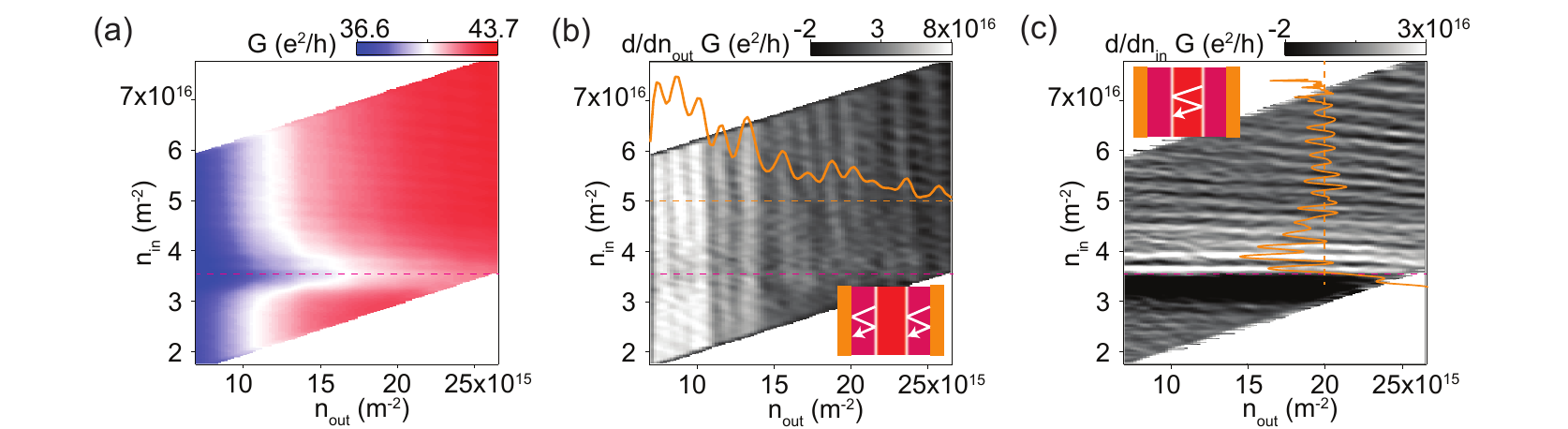}
    \caption{\textbf{FP resonances in the inner and outer cavities across the satellite DP. a,} Conduction as a function of the charge-carrier density in the inner and outer cavity. The position of the satellite DP is indicated with the dashed, purple line. Numerical derivative along \textbf{b,} $n_{\text{out}}$ and \textbf{c,} $n_{\text{in}}$ reveal the conduction resonances in the outer and inner cavity respectively. Two representative line profiles of the numerical derivative are shown in orange.}
    \label{fig:FP_Moiree}
\end{figure}
Compared to the FP resonances observed across the main DP (bipolar region), the visibility for the resonances in the outer cavities is less pronounced compared to the one in the inner cavity. Extracting the cavity-length from the linecuts indicated in Fig.~\ref{fig:FP_Moiree}b and \ref{fig:FP_Moiree}c confirms that the observed FP resonances belong to resonances in the outer and inner cavity respectively.

\subsection{Energy scales in the Moir\'{e} mini-bands and residual doping}
\begin{figure}[htbp]
    \centering
      \includegraphics[width=1\textwidth]{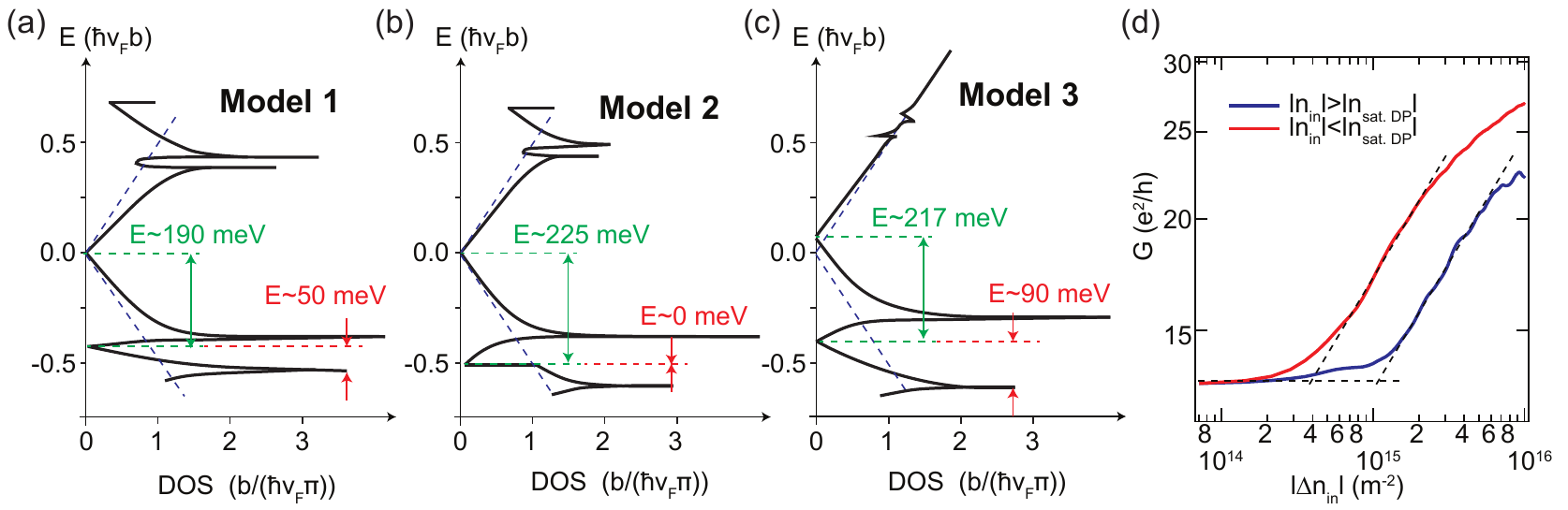}
    \caption{\textbf{Energy scales within the Moir\'{e} mini-bands and residual doping. a-c,}  Calculated energy spacing between main and satellite DP (green) and the satellite DP and the first Van Hove singularity at higher energy (red) for a rotation angle of $\theta=0.85 ^{\circ}$ between graphene and hBN (b,c and d corresponds to the same theoretical model with different parameters). Figure adapted from Ref.~\cite{Wallbank13}. \textbf{d,} Residual doping in the system extracted from the satellite DP.}
    \label{fig:Theory_Bandstructure}
\end{figure}

The purpose of this section is to verify that the energy smearing due to residual doping in our system is smaller than the energy range within which only the first Moir\'{e} mini-band for $E<0$ is populated.

We start by comparing known values from the experiment with those extracted from theoretical calculations, namely the energy-spacing between the main and satellite DP (hole doping), indicated with green dashed lines in Fig.~\ref{fig:Theory_Bandstructure}a-c. In the experiment, the latter is given by
\begin{equation}
\Delta E=\hbar v_{\text{F}} \sqrt{\pi n_{\textrm{sat.DP}}}
\end{equation}
where $n_{\textrm{sat.DP}}$ is the position of the satellite DP with respect to the main DP. Using $n_{\textrm{sat.DP}} \sim 3.67\times 10^{16}\,$m$^{-2}$ extracted from the experiment, one finds $E\sim225\,$meV. This number can be compared with the values extracted from the theoretical calculations shown in Fig.~\ref{fig:Theory_Bandstructure}a-c (green dashed lines), where the energy is plotted in units of $\hbar v_{\text{F}} b$, where $b$ is a reciprocal lattice spacing. The latter is given by $b=\frac{4 \pi}{3 a}\sqrt{\delta ^2+\theta ^2}$, where $a=0.142\,$nm is the distance between neighbouring carbon atoms in graphene, $\delta=$0.018 is the lattice mismatch between graphene and hBN and $\theta$ is the rotation angle between graphene and hBN. This rotation angle can be extracted as well from $n_{\textrm{sat.DP}}$ as described in Ref.~\cite{Ponomarenko13,Yankowitz12}, being $\theta\sim0.85\, ^\circ$ in our system. Finally it can be seen that the theoretical values, ranging from $E=190-225\,$meV, are in good agreement with the experimental values.

Depending on the parameters used in the theoretical model, the band structure of the Moir\'{e} mini-bands varies quite significantly, as shown in Fig.~\ref{fig:Theory_Bandstructure}a-c. The energy range within only the first Moir\'{e} mini-band ($E<0$) is populated is roughly given by the spacing between the satellite DP (DOS vanishes) and the first Van Hove singularity of the DOS at higher energy. This energy spacing, indicated with the red dashed lines in Fig.~\ref{fig:Theory_Bandstructure}a-c, is ranging between zero and $90\,$meV.

We compare the theoretically extracted values with the energy smearing extracted from the residual doping in the system as shown in Fig.~\ref{fig:Theory_Bandstructure}d. The residual doping was extracted from the satellite DP, but comparable values result from a similar analysis of the main DP. Note that in Fig.~\ref{fig:Theory_Bandstructure}d zero density is set at the minimal conductance of the satellite DP, thus $\Delta n_{\text{in}}=n_{\text{in}}\text{(main DP)}-n_{\text{in}}\text{(sat.DP)}$. The residual doping can be converted into an energy according to
\begin{equation}\label{Energy_SDP}
E=\hbar v_{\text{F,sat.DP}} \sqrt{\frac{\pi n_{\text{res}}}{g_{\text{SDP}}}}
\end{equation}
where $v_{\text{sat.DP}}=0.45\cdot 10^{6}$ $m/s$ is the rescaled Fermi velocity \cite{Yu14}
at the satellite DP and $g_{\text{SDP}}$ is the additional degeneracy of the satellite Dirac spectra. Following model 3 ($g_{\text{SDP}}=1$), and assuming a constant Fermi velocity within the given doping range (strictly speaking this is only valid in the very proximity of the satellite DP), the residual doping corresponds to an energy of $E\sim 15\,$meV on the hole side. Considering model 3 the residual doping does not seem to be the limiting factor in resolving features within the first Moir\'{e} mini-band, unless these features are in the very close vicinity of the  satellite DP. For consistency, we can convert the energy spacing of $90\,$meV, indicated with the red dashed lines in Fig.~\ref{fig:Theory_Bandstructure}c into a charge carrier doping, leading to $\Delta n=3 \cdot 10^{16}\;$m$^{-2}$.

\subsection{Residual doping and mode index}
The residual doping ($n_{\text{res}}$) defined by the puddle landscape, can be converted into energy according to
\begin{equation}
E_{\text{res}}=\hbar v_{\text{F}} \sqrt{\pi n_{\text{res}}}.
\end{equation}
 In the device without Moir\'{e} superlattice the residual doping was of order $n_{\text{res}}\sim2\text{-}6\cdot 10^{14}\,$m$^{-2}$. On the other hand, the energy of the FP resonances is given by
\begin{equation}
E_{\text{FP}}=\hbar v_{\text{F}} j \frac{\pi}{L}
\end{equation}
where $j$ is the mode-index ($j=1,2,3...$) coming from the constructive resonance condition $L=j\cdot \lambda/2$. Fabry-P\'{e}rot resonances with a smaller energy-spacing than the puddle landscape cannot be resolved, consequently $E_{\text{FP}}\ge E_{\text{res}}$, or
\begin{equation}
j\ge L\sqrt{\frac{n_{\text{res}}}{\pi}}.
\end{equation}
For $L\sim200\,$nm (at low doping as shown in the main text), this leads to $i\ge3$ which is in good agreement with the experiment ($j= L\sqrt{n/\pi}\sim 3$ with $n\sim5\cdot 10^{14}\,$m$^{-2}$ for the lowest observed FP oscillation).

\bibliographystyle{PRL}

\end{document}